\documentstyle[twocolumn,aps,epsf]{revtex}
\begin{document}
\title{Analog of Magnetoelectric Effect in High-$T_c$ Granular
Superconductors}
\author{Sergei A. Sergeenkov$^{1}$\cite{byline} and Jorge V. Jos\' e$^{2,3}$}
\address{$^{1}$Instituto de Fisica, Universidade Federal do Rio Grande do Sul,
91501-970 Porto Alegre, Brazil\\
$^{2}$Physics Department and Center for the Interdisciplinary
Research on Complex Systems \\ Northeastern University, Boston, MA,
02115, USA \\ $^{3}$Instituto de F\'{\i}sica,
Universidad Nacional Aut\' onoma de M\' exico,
Apdo. Postal 20-364, 01000 M\' exico D. F., M\' exico}
\address{~}
\address{
\centering{
\begin{minipage}{16cm}
We propose the existence of an  electric-field induced
nonlinear magnetization in a weakly coupled granular superconductor
due to time-parity violation.
As the field increases the induced magnetization passes from para- to
dia-magnetic behavior. We discuss  conditions under which this effect
could be experimentally measured in high temperature superconductors.
\end{minipage}
}}
\maketitle
\narrowtext
By applying a constant external electric field through an insulating
layer, recent experimental studies~\cite{ref1,ref2,ref3} have found
an {\it enhancement} in the critical current of
ceramic high-$T_c$ superconductors (HTCS).
It has been shown that for any field-induced effects to be
experimentally measurable, the electrostatic screening length
should be larger than the superconducting coherence length.
This happens to be the case in HTCS, which have a reduced carrier number
density~\cite{ref4,ref5}. To account for the unusual behavior
of the field dependent critical current observed in HTCS
ceramics~\cite{ref1,ref2,ref3}, however, it was argued that
some other mechanisms (in addition to the conventional carrier
number density modification) have
to be taken into consideration. In particular, it was suggested that an
applied electric field may actually induce a substantial change into the
original weak-links structure of granular samples~\cite{rak}.

In this letter we predict yet another interesting effect which can be produced
by an external electric field applied to a granular material. This effect is
similar to (but physically different from) the {\it magnetoelectric effect}
(MEE) seen in magnetic ferroelectrics (like antiferromagnetic
$BiFeO_3$)~\cite{ref5b}. It has also been discussed in the context of some
exotic nonmagnetic normal metal conductors,  with the symmetry of a mirror
isomer~\cite{levitov}, and pyroelectric superconductors (where the
supercurrent passing through a metal of a polar symmetry is assumed to
be accompanied by spin polarization of the carriers~\cite{SO}).
The effect discussed here entails the
electric field generation of a magnetic moment (paramagnetic in small
 to diamagnetic for large electric fields) due to
superconducting currents that circulate between grains.
The basic physical reason for the appearance of this effect is that the
applied electric field induces a magnetization $M$, that changes its sign
($M\to -M$) under a time-parity transformation. There have been other
microscopic mechanisms proposed to have time-parity violation effects in
HTCS. One that attracted significant experimental attention was a
theoretical consequence of the anyon picture of HTCS~\cite{anyon}.
Another proposition was based on a spin-orbit coupling~\cite{SO}.
Up to now, however, the experimental search for their signatures
have not bared fruit.

In the case of weakly-coupled superconducting grains, the
 phenomenological reason for the time-parity violation comes
from the fact that the total free energy $F$ for the
material exhibiting both magnetic and electric properties
contains the term $\alpha_{mn}f_{mn}(\vec E,\vec H)$,
with the coefficients $\alpha_{mn}\neq 0$ (here $\{m,n\}=x,y,z$).
In the standard MEE, the function $f_{mn}\equiv E_mH_n$, which leads to
the corresponding {\it linear} effect, which is either an electric-field
induced magnetization $M_m(\vec E)\equiv \partial F/\partial H_m=\alpha
_{mn}E_n$ (in zero magnetic field) or a magnetic-field induced polarization
$P_m(\vec H)\equiv \partial F/\partial E_m=\alpha _{mn}H_n$ (in zero
electric field). Since $H$ (and $M$) changes its sign under time-parity
transformation while $E$ (and $P$) remains unchanged, an electric field
induced magnetization will break time-parity symmetry even in zero magnetic
field applied.
As we show below, in our case the symmetry breaking term in $F$,
represented by a nonzero coefficient $\alpha_{mn}$, has a
more general {\it nonlinear} form for $f_{mn}$.
We recall that the standard linear MEE can appear when an external electric
field $\vec E$ interacts with an inner magnetic field $\vec h_{DM}$ of the
Dzyaloshinskii-Moriya (DM) type~\cite{moriya}. The DM interaction
leads to a term, apart from the standard isotropic term
 in the Heisenberg Hamiltonian, of the form
$H_{DM}= \sum _{i,j} \vec D_{i,j}\cdot (\vec S_i\wedge
\vec S_j)$, where $\vec S_i$ is a Heisenberg spin and the constant
vector $\vec  D_{i,j}$ arises from the spin-orbit coupling.
An analogous situation occurs in our case, as we describe below.
To see how we can get a nonzero $\alpha$, or equivalently a DM type
interaction in a granular superconductor, we model a HTCS ceramic sample by a
{\it random} three-dimensional (3D) overdamped Josephson junction array.
This model has proven to be useful in describing the
metastable magnetic properties of HTCS~\cite{ebner,choi}. In thermodynamic
equilibrium, this model has a Boltzmann factor with a random 3D-XY model
Hamiltonian.
Specifically, the general form of the Hamiltonian (describing
both DC and AC effects) reads
\begin{equation}
{\cal H}_{XY}(t)=-\sum_{i,j}E_J(r_{i,j}) \cos \phi_{i,j}(t).
\end{equation}
Here $\{i\}=\vec {r}_i$ is a 3D lattice vector; $E_J(r_{i,j})$ is the Josephson
coupling energy, with $\vec r_{i,j}=\vec r_i-\vec r_j$ the separation
between the grains;
the  gauge invariant phase difference is defined as
\begin{equation}
\phi _{i,j}(t)=\phi _{i,j}(0)-A_{i,j}(t),
\end{equation}
where $\phi _{i,j}(0)=\phi _i-\phi _j$ with $\phi_i$ being the phase of the
superconducting order parameter; $A_{i,j}(t)$ is (time-dependent, in general)
frustration parameter, defined as
\begin{equation}
A_{i,j}(t)=\frac{2\pi}{\Phi_ 0}\int_i^j\vec A(\vec r,t)\cdot d{\vec l},
\end{equation}
with $\vec A(\vec r,t)$ the (space-time dependent) electromagnetic vector
potential which involves both external fields and the electric and
magnetic possible  self-field
effects (see below); $\Phi_ 0=h/2e$ is the quantum of flux, with $h$
Planck's constant, and $e$ the electronic charge.
Expanding the cosine term, and using trigonometric identities
we can explicitly rewrite the above Hamiltonian as
$H_{XY}=-\sum _{i,j}E_J[\cos (A_{i,j})\vec S_i\cdot\vec S_j-\sin (A_{i,j})
{\hat k}\cdot {\vec S_i}\wedge {\vec S_j}]$,
where the two-component XY spin vector is defined as $\vec S_i\equiv
(\cos\phi_i, \sin\phi_i)$, and $\hat k$ is a unit vector along the
z-axis \cite{gingr}.
We see that the second term in this Hamiltonian has the
same form as in the DM contribution, and thus we can surmise that
the time parity will be broken by applying an external field (electric or
magnetic) to the granular system. To bring to the fore this possibility, we
show below in a simple but yet nontrivial model that this is indeed the case.

One important property of the $H_{XY}$ Hamiltonian is that it is
random because $A_{i,j}$ itself is a random variable.
There are
different types of $A_{i,j}$ randomness that can be considered~\cite{choi}.
For simplicity, in the present paper, we consider a long-range
(Sherrington-Kirkpatrick-like) interaction between grains
(assuming $E_J(r_{i,j})=E_J$) and model the true short-range behavior
of a ceramics sample through the randomness in the
position of the superconducting grains in the array (using the exponential
distribution law $P_r(r_{i,j})$, see below).
Here we restrict our consideration
to the case of an external electric field only but it can be shown that the
scenario suggested in this paper will also carry through when applying an
external magnetic field (which will induce another time-parity breaking
phenomenon in the granular material; namely, magnetic field induced electric
polarizability~\cite{ser}). Besides, in what follows we also ignore
the role of Coulomb interaction effects assuming that the grain's charging
energy $E_c\ll E_J$ (where $E_c=e^2/2C$, with C the capacitance of
the junction).

In the case of a granular material, we show here that
the corresponding time-parity breaking DM
internal field can be related to the electric field induced
magnetic moment produced by the
circulating Josephson currents between the grains.
As is known~\cite{tinkham,lebeau}, a constant electric field $\vec E$ applied
to a single Josephson junction (JJ) causes a time evolution of the phase
difference. In this particular case Eq.(2) reads
\begin{equation}
\phi _{i,j}(t)=\phi _{i,j}(0)+\frac{2e}{\hbar}\vec E\cdot \vec r_{i,j}t.
\end{equation}
The resulting  AC superconducting current in the junction is
\begin{equation}
I_{i,j}^s(t)=\frac{2eE_J}{h} \sin \phi _{i,j}(t).
\end{equation}
If, in addition to the external electric field $\vec E$, the network of
superconducting grains is under the influence of an applied magnetic field
$\vec H$, the frustration parameter $A_{i,j}(t)$ in Eq.(3) takes the
following form
\begin{equation}
A_{i,j}(t)=\frac{\pi}{\Phi _0}(\vec H\wedge \vec R_{i,j})\cdot \vec r_{i,j}-
\frac{2\pi}{\Phi _0}\vec E\cdot \vec r_{i,j}t.
\end{equation}
Here, $\vec R_{i,j}=(\vec r_i+\vec r_j)/2$, and we have used the conventional
relationship between the vector potential $\vec A$ and a constant
magnetic field $\vec H=rot \vec A$ (with $\partial \vec H/\partial t=0$), as
well as a homogeneous electric field $\vec E=-\partial \vec A/\partial t$ (with
$rot \vec E=0$). In the type II  HTCS the magnetic
self-field effects for the array as a whole are expected to be
negligible~\cite{dom}. The grains themselves are in fact larger than the
London penetration depth and we must then have that the
corresponding Josephson penetration length must be much larger than the
grain size
(since the self-induced magnetic fields can in principle
be quite pronounced for large-size junctions even in zero applied magnetic
fields~\cite{ser}.)
Specifically, this  is justified for short junctions with
the size $d\ll \lambda _J$, where $\lambda _J=\sqrt{\Phi _0/4\pi \mu _0j_c
\lambda _L}$ is the Josephson penetration length  with $\lambda _L$ being
the grain London penetration depth and $j_c$ its Josephson critical current
density. In particular, since in HTCS
$\lambda _L\simeq 150nm$, the above
condition will be fulfilled for $d\simeq 1\mu m$ and $j_c\simeq 10^{4}A/m^2$
which are the typical parameters for HTCS ceramics~\cite{tinkham}.
Likewise, to ensure the uniformity of the applied electric field, we also
assume that $d\ll \lambda _E$, where $\lambda _E$ is an effective electric
field penetration depth~\cite{rak}.

When the AC supercurrent $I_{i,j}^s(t)$ (defined by Eqs.(2), (5) and (6))
circulates around a set of grains,
that form a random area plaquette, it induces a random AC magnetic
moment $\vec \mu(t)$ of the Josephson network~\cite{ebner}
\begin{equation}
\vec \mu (t)\equiv \left [\frac{\partial
{\cal H}_{XY}}{\partial \vec H}
\right ]_{\vec H=0}=
\pi \sum_{i,j}I_{i,j}^s(t)(\vec r_{i,j}\wedge \vec R_{i,j}).
\end{equation}
Notice that
in the MEE-like effect discussed here for a granular superconductor,
the electric-field induced magnetic moment in the
system is still present in zero applied magnetic field due to the
phase coherent currents between the weakly-coupled superconducting grains.

To consider the essence of the superconducting electric field-induced MEE,
we assume that in a {\it zero electric field} the phase difference
between the adjacent grains $\phi _{i,j}(0)=0$ which corresponds to
a fully coherent state of the array. In this particular case, the
electric-field induced averaged magnetization reads
 \begin{equation}
\vec M_s(\vec E)\equiv \overline{\vec \mu_s (t)}=
\frac{1}{\tau}\int\limits_{0}^
 {\tau }dt \int\limits_{0}^{\infty }d\vec r_{i,j}d\vec R_{i,j}
S(\vec r_{i,j}, \vec R_{i,j}) \vec \mu (t),
 \end{equation}
where $\tau$ is the electronic relaxation scattering time,
and $S$ is the joint probability distribution function (see below).

To obtain an explicit expression for the electric-field dependent
magnetization, we consider a site positional disorder that allows for small
random radial displacements. Namely, the sites in a 3D cubic lattice are
assumed to move from their equilibrium positions according to the normalized
(separable) distribution function
\begin{equation}
 S(\vec r_{i,j}\vec R_{i,j})\equiv P_{r}(\vec r_{i,j})P_{R}(\vec R_{i,j})
\end{equation}
It can be shown that the main qualitative results of this paper do not depend
on the particular choice of the probability distribution function.
For simplicity here we assume an exponential distribution law for the
distance between grains, $P_r(\vec r)=P(x_1)P(x_2)P(x_3)$ with
$P_r(x_j)=(1/d)e^{-x_j/d}$, and
a short range distribution for the
dependence of the center-of-mass probability $P_R(\vec R)$ (around
some constant value $D$). The specific form of the latter distribution
will not affect the qualitative nature of the final result.
(Notice that in fact the former
distribution function $P_r(\vec r)$ reflects a short-range character of the
Josephson coupling in granular superconductor. Indeed, according to the
conventional picture~\cite{comment}
the Josephson coupling $E_J(\vec r_{ij})$ can be  assumed to vary
exponentially with the distance $\vec r_{ij}$ between neighboring grains,
i.e., $E_J(\vec r_{ij})=E_Je^{-\vec \kappa \cdot \vec r_{ij}}$.
For isotropic arrangement of identical grains, with spacing
$d$ between the centers of adjacent grains, we have
$\vec \kappa =(\frac{1}{d},\frac{1}{d},\frac{1}{d})$ and thus $d$ is of the
order of an average grain size.)
Taking the applied electric field along the
$x$-axis, $\vec E=(E_x,0,0)$, we get finally
\begin{equation}
 M_z(E_x)=\frac{B_z(E_x)}{\mu _0}-H_z(E_x),
 \end{equation}
for the induced transverse magnetization (along the $x_3=z$-axis), where
 \begin{equation}
 B_z(E_x)=\mu _0M_0\frac{E_x/E_0}{1+(E_x/E_0)^2},
 \end{equation}
 and
 \begin{equation}
 H_z(E_x)=M_0\left (\frac{E_0}{E_x}\right ) \log \sqrt{1+
\left (\frac{E_x}{E_0}\right )^2},
 \end{equation}
stand for the electric-field induced  magnetic induction $B_z(E_x)$
and magnetic field $H_z(E_x)$, respectively. The induced
Josephson current $I(E_x)$ is simply given by Ampere's law $I(E_x)=H_z(E_x)d$.
In these equations, $M_0=2\pi eE_JNdD/\hbar $, with $N$ the total number of
grains and
$E_0=\hbar /2de\tau$. Eq.(10) is the main result of this paper, which we
proceed to analyze below.
\begin{figure}
\epsfxsize=8.5cm
\centerline{\epsffile{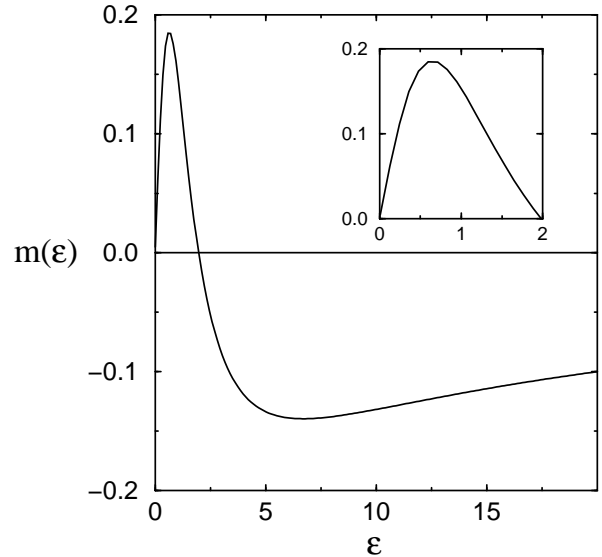} }
\caption{The induced magnetization $m(\epsilon)=M_z/M_0$ as
a function of the normalized applied electric field $\epsilon =E_x/E_0$,
according to Eq.(10). The inset shows a blow up of the paramagnetic
region of $m(\epsilon)$ (exhibiting its linear superconducting
magnetoelectric effect regime). }
\end{figure}
As is seen from Eq.(10), the behavior of the magnetization in the applied
electric field is determined by the competition between the two
contributions, the magnetic induction $B_z(E_x)$ and the current induced
magnetic field $H_z(E_x)$ (or the corresponding Josephson current $I(E_x)$).
Namely, below a critical (threshold) field $E_c\approx 1.94\, E_0$ where
$B_z(E_x)>\mu _0H_z(E_x)$, a paramagnetic phase of the MEE is due to
the modification of the magnetic induction in the applied electric field.
On the other hand above the threshold field (when $E_x$ becomes larger
than $E_c$) the Josephson current $I(E_x)$ induced contribution
starts to prevail, leading to the appearance of the diamagnetic
signal (seen as a small negative part  of the induced magnetization
in Fig.1). Such an electric field induced
paramagnetic-to-diamagnetic transition has been actually
observed~\cite{ref1,ref2,ref3} for behavior of the critical current in
ceramic HTCS and it was attributed~\cite{rak} to a proximity-mediated
enhancement of the superconductivity in a granular material in a
strong enough electric field. Explicitly, the critical current in a $YBCO$
sample was found to reach a maximum  at $E=4\times 10^7 V/m$.
To relate
this experimental value with the model parameters,
first of all, we need to estimate an order of magnitude of the relaxation
time $\tau$ in a zero applied electric field. This will provide an {\it upper}
limit for the relevant
$\tau$-distribution in our system. It is reasonable to connect zero-field
$\tau \equiv \tau (0)$ with the Josephson tunneling time~\cite{tinkham}
$\tau _J =(R_0/R_n)(\hbar /E_J)$ (where
$R_0=h/4e^2$, and $R_n$ is the normal state resistance between grains).
Typically, for HTCS ceramics $E_J/k_B\simeq 90K$ and
$R_n/R_0\simeq 10^{-3}$, so that $\tau _J\simeq 10^{-10}s$.
At the same time, at high enough electric
fields where the MEE becomes strongly nonlinear, we can expect quite a
tangible decrease of the relaxation time.
Indeed, for an average grain size $d\approx 1\mu m$, the characteristic
field $E_0=4\times 10^7 V/m$ (which corresponds to the region where a
prominent enhancement of the critical current
was observed~\cite{ref1,ref2,ref3}) introduces a substantially shorter
relaxation time $\tau (E_0)=\hbar /2deE_0\simeq 10^{-16}s$, in agreement
with observations~\cite{ref4}.

To estimate the relative magnitude of the superconducting analog of the MEE
predicted here, we can compare it with the normal (Ohmic) contribution to
the magnetization
\begin{equation}
\vec M_n(\vec E)\equiv \overline{\vec \mu_n (t)}=\frac{1}{\tau}\int\limits_{0}^
{\tau }dt \int\limits_{0}^{\infty }d\vec r_{i,j}d\vec R_{i,j}S(\vec r_{i,j},
\vec R_{i,j}) \vec \mu _n,
\end{equation}
where
\begin{equation}
\vec \mu _n=\pi \sum_{i,j}I_{i,j}^n(\vec r_{i,j}\wedge  \vec R_{i,j}).
\end{equation}
Here $I_{i,j}^n=V_{i,j}/R_n$ is the normal current component  due to
the applied electric field $\vec E$, with $V_{i,j}=\vec E\cdot \vec r_{i,j}$
being the induced voltage, and $R_n$ the normal state resistance between
grains. As a result, the normal state contribution (for $\vec E$ along the
$x$-axis) reads $M_n=\alpha _nE_x$, with $\alpha _n=\pi d^2DN/R_n$.
Similarly, according to Eq.(10), the low field contribution to the
superconducting MEE gives $M_s\simeq \alpha _sE_x$ with
$\alpha _s=2\pi e^2E_JN\tau (0)d^2D/\hbar ^2$.
Thus, at low enough applied fields (when $E_x\ll E_0$)
\begin{equation}
\frac{\alpha _s}{\alpha _n}\simeq \frac{\tau (0)}{\tau _J},
\end{equation}
where $\tau _J =(R_0/R_n)(\hbar /E_J)$ with $R_0=h/4e^2$.
According to our previous discussion on the relevant relaxation-time
distribution spectrum in our model system, we may conclude that
$\tau (0)\leq \tau _J$. So, we arrive at the
following ratio between the coefficients of the superconducting to normal
MEEs, namely $\alpha _s/\alpha _n\leq 1$. The above estimate of the
weak-links induced MEE (along with its rather specific field
dependence, see Fig.1) suggests quite an optimistic possibility to observe
the predicted effect experimentally in HTCS ceramics or in a specially
prepared system of arrays of superconducting grains.

We note that in the present analysis we have not
explicitly considered the polarization effects (due to the interaction
between the applied electric field and the grain's charges) which may become
important at high enough fields (or for small enough grains), leading to
more subtle phenomena (like Coulomb blockade and reentrant-like behavior)
that will demand the inclusion of charging energy effects in the analysis.

In summary, we have used a zero-temperature random 3D XY model to predict the
appearance of a novel electric-field induced magnetization in a granular
superconductor (an analog of the magnetoelectric effect). The induced
magnetization has a very distinctive nonlinear and para- to dia-magnetic
behavior as the field is increased.  We have also
estimated the possible parameter range conditions to observe this effect
experimentally.

We thank N. Israeloff, A. Goldman and P. Martinoli for informative
communications on the
subject. SAS acknowledges the financial support from the Brazilian funding
agency  CNPq. The work of JVJ was partially funded by NSF Grant DMR-9521845.

\end{document}